\documentclass[aps,prd,floatfix,noshowpacs,twocolumn,10pt,showkeys]{revtex4-1}

\usepackage{siunitx}

\usepackage{amsmath}
\usepackage{amssymb}

\usepackage{amsfonts}
\usepackage{mathrsfs}

\usepackage{graphicx}
\usepackage{dcolumn}
\usepackage{bm}
\usepackage{epsfig}
          \usepackage{epstopdf}
\usepackage{enumerate}
 \usepackage{float}
\usepackage{amssymb}

\usepackage{color}

\usepackage[colorlinks=true,linkcolor=darkblue,citecolor=darkblue,urlcolor=darkblue,breaklinks=true]{hyperref}

\usepackage[hyphenbreaks]{breakurl}
\definecolor{darkblue}{rgb}{0,0,0.7}

\def\be{\begin{equation}}
\def\ee{\end{equation}}
\def\bea{\begin{eqnarray}}
\def\eea{\end{eqnarray}}
\def\bfl{\begin{flushleft}}
\def\efl{\end{flushleft}}
\def\bfr{\begin{flushright}}
\def\efr{\end{flushright}}
\def\bc{\begin{center}}
\def\ec{\end{center}}
\def\ben{\begin{enumerate}}
\def\een{\end{enumerate}}
\def\bit{\begin{itemize}}
\def\eit{\end{itemize}}

\def\dzn{,\kern-0.1em,}

\def\d#1{{#1\kern-0.4em\char"16\kern-0.1em}}
\def\D#1{{\raise0.2ex\hbox{-}\kern-0.4em 31}}

\def\d{\text{d}}

\newcommand {\apgt} {\ {\raise-.5ex\hbox{$\buildrel>\over\sim$}}\ }
\newcommand {\aplt} {\ {\raise-.5ex\hbox{$\buildrel<\over\sim$}}\ }

\begin{document}




\title{Exciton dynamics in different aromatic hydrocarbon systems}


\author{Milica Rutonjski}
\affiliation{Department of Physics, Faculty of Sciences, University of Novi Sad, Trg Dositeja
 Obradovi\' ca 4, Novi Sad, Serbia}
\email{milica.rutonjski@df.uns.ac.rs}
\author{Petar  Mali}
\affiliation{Department of Physics, Faculty of Sciences, University of Novi Sad, Trg Dositeja
 Obradovi\' ca 4, Novi Sad, Serbia}
\author{Slobodan  Rado\v sevi\' c}
\affiliation{Department of Physics, Faculty of Sciences, University of Novi Sad, Trg Dositeja
 Obradovi\' ca 4, Novi Sad, Serbia}
\author{Sonja Gombar}
\affiliation{Department of Physics, Faculty of Sciences, University of Novi Sad, Trg Dositeja
 Obradovi\' ca 4, Novi Sad, Serbia}
\author{Milan Panti\' c}
\affiliation{Department of Physics, Faculty of Sciences, University of Novi Sad, Trg Dositeja
 Obradovi\' ca 4, Novi Sad, Serbia}
 \author{Milica Pavkov-Hrvojevi\' c}
\affiliation{Department of Physics, Faculty of Sciences, University of Novi Sad, Trg Dositeja
 Obradovi\' ca 4, Novi Sad, Serbia}






\begin{abstract}
The exciton dispersion is examined in the case of four selected prototypical molecular solids: pentacene,tetracene,picene,chrysene. 
The model parameters are determined by fitting 
to experimental data obtained by inelastic electron 
scattering. Within the picture that relies on Frenkel-type excitons we obtain that theoretical dispersion curves along different directions in 
the Brillouin zone are in good agreement with the experimental data, suggesting that the influence of charge-transfer excitons on exciton dispersion of the analyzed organic solids is not as large as proposed. In reciprocal space directions where Davydov splitting is observed we employ the upgraded version of Hamiltonian used in \href{https://www.mdpi.com/1996-1944/11/11/2219}{Materials \textbf{11}, 2219 (2018)}.

\end{abstract}





\maketitle

\section {Introduction}
Organic semiconductors have been in the focus of both theoretical and experimental studies for decades. This comprehensive research has been motivated 
by the wide field of their application in novel (opto)electronic devices \cite{tranzistor,naturePeumans,natureBaldo,forest,jang}. Besides, it has been observed that some of them, picene for example, exhibit the transition into the superconducting state at rather high transition temperatures \cite{nature}. 
Therefore, the microscopic properties of aromatic hydrocarbons, in particular their exciton dynamics, present the subject of great interest.

Electron energy-loss spectroscopy is an experimental technique that has been recently widely used for
direct measurements of the exciton band structure in aromatic hydrocarbons \cite{183,breakdown,Ugaopicen,rothpentacensvi,rothlepaslika,rothtetracen,rothkrizen}. 
These experiments inspired significant theoretical work based on the first principles, i.e. starting from many-body electron-hole Hamiltonians  \cite{GatuzoPRB,GatuzoPRB2,GatuzoJPC,AnnRev,JPCLett}. 
In these papers it is suggested that exciton dispersion in the organic molecular solids known as phenacenes (picene, chrysene) can be understood within the Frenkel-exciton picture, whereas the contribution of charge-transfer (CT) excitons in the lowest-lying exciton states in the so-called acenes (pentacene, tetracene) is more significant.
However, in our recent paper \cite{Materials} where CT excitons are completely ignored, the pentacene exciton dispersion in a good agreement with the experimental data was obtained by making use of the effective anisotropic Heisenberg model in external field in Bloch approximation. Unlike previous theoretical works based on many-body Hamiltonians containing electron and hole creation and annihilation operators, the effective Hamiltonian from \cite{Materials} contains exciton ladder operators. Namely, starting from 
Frenkel excitons as low-lying degrees of freedom, the effects of their interactions to the one-loop order is shown to be negligible in wide temperature interval. 
However, due to the limited application of the model, we investigated only those Brillouin zone directions where Davydov splitting \cite{davydov} was not observed. Therefore, in the present paper the model is generalized in order to reproduce the experimental data for both Davydov components. The subject of study is expanded to other experimentally studied aromatic hydrocarbons, such as tetracene, picene and chrysene, enabling us to compare the excitation spectra of the acenes family to the ones of phenacenes, which are characterized by a relatively large band gap.

The paper is organized as follows. The model Hamiltonian
and the crystal structures of all analyzed molecular solids are introduced
in Sec. II. Exciton dispersions along different directions within reciprocal $\bm a^*\bm b^*$ plane
are obtained, discussed and compared to existing experimental data in Sec. III. Finally, the conclusions are drawn in Sec. IV.

\section {Model Hamiltonian} 

Within the picture that relies on Frenkel-type excitons in two-level systems and their predominant nearest neighbour interaction,  
the basic Hamiltonian is given by
\bea 
H&=&H_0+\Delta \sum_{\bm{n}}P^+_{\bm{n}}P_{\bm{n}}
-\frac{X}{2}\sum_{\bm{n},\bm{\lambda}}P^+_{\bm{n}}P_{\bm{n}+
\bm{\lambda}}\nonumber\\
&-&\frac{Y}{2}\sum_{\bm{n},\bm{\lambda}}P^+_{\bm{n}}P_{\bm{n}}
P^+_{\bm{n}+\bm{\lambda}}P_{\bm{n}+\bm{\lambda}}, \label{paulham} 
\eea
where $P^+_{\bm{n}}$ and $P_{\bm{n}}$ represent standard Pauli 
operators on the site $\bm{n}$, whereas parameters $X$ and $Y$ respectively describe
hopping and interactions of excitons \cite{Agranovicbook,Tole}. In a two-level system it is possible to find exact mapping  
between Pauli Hamiltonian (\ref{paulham}) and anisotropic 
($XXZ$) Heisenberg Hamiltonian in external field \cite{Materials}
\be 
H=-\frac{I^x}{2}\sum_{\bm{n},\bm{\lambda}}S^-_{\bm{n}}
S^+_{\bm{n}+\bm{\lambda}}-\frac{I^z}{2}\sum_{\bm{n},\bm{\lambda}}
S^z_{\bm{n}}S^z_{\bm{n}+\bm{\lambda}}-\mu \mathcal{H}
\sum_{\bm{n}}S^z_{\bm{n}}. \label{spinham}
\ee
Parameters $I^{x/z}$ are the exchange integral components, $\mathcal{H}$ represents external field, while vectors $\{\bm{\lambda}\}$ connect nearest neighboring sites.
Correspondence between model parameters 
\bea
I^z = Y, \quad
I^x = X, \quad
\mu \mathcal{H} = \Delta - \frac{I^z z_1}{2} \label{veza}
\eea
is justified due to the isomorphism of paulion Hilbert space $\mathscr{H}_{\rm P}$ and spin Hilbert space $\mathscr{H}_{\rm S}$.
This procedure, which is purely theoretical, is motivated by the fact that the literature referring to spin systems and the theoretical tools therein developed are significantly richer \cite{Tjablikov,Kuntz,Averbah,Mano,Nolting,Sandvik,SSC,Hofman1,Hofman2}.

The direct application of Agranovich-Toshich representation \cite{Tole} on Hamiltonian (\ref{paulham}) would lead to unnecessary complications: the Hamiltonian expressed in terms of Pauli operators already contains fourth-order terms and therefore the effects of exciton-exciton interactions are split into three parts. The first one comes from the Pauli operator commutation relations, the second one from nontrivial Hilbert space and the third one from direct fourth-order Pauli operator terms. The theory based on spin Hamiltonian (\ref{spinham}) is simpler since it lacks the issue of fourth order spin-operator terms. However, it is still flawed by first two problems. There is an alternative approach, which dismisses boson representations altogether. It is the method of effective Lagrangians and arguments presented here (as well as in related paper \cite{Materials}) are based upon it. In the method of effective Lagrangians, one directly uses physical degrees of freedom. The effective Lagrangian already contains exciton operators and their interactions -- they do not emerge via boson representation. Also, a significant feature of these interactions is their manifest weakness which allows for a systematic organization of Feynman diagrams (details can be found in \cite{Materials,radosevic2013,radosevic2015}. The transition to effective Lagrangian method (in this case) requires a Hamiltonian which can be split into O(3) invariant part and a symmetry breaking term. Therefore, the use of Heisenberg Hamiltonian in an intermediate step towards an effective model is mandatory. Our previous paper \cite{Materials} demonstrated that exciton-exciton interactions in such effective model are indeed negligible. Therefore, the free boson Hamiltonian used in this paper should not be viewed as obtained with some boson representation of Pauli/spin operators but merely as an interaction-free term dictated by the effective Lagrangian.

Note also that, in case of aromatic hydrocarbons analyzed in this paper, the set of neighboring sites connected with 
hopping integrals splits into three subsets determined by the 
lattice structure and values of hopping parameters. 

We shall now focus on the structure of the polycyclic aromatic hydrocarbon systems studied in this paper. Due to the molecular structure differences, tetracene and pentacene belong to the so-called acenes family, while picene and chrysene represent the examples of phenacenes. The former two crystallize in triclinic and the latter two in monoclinic crystal system. However, since the carrier mobility along $\bm {c}^*$ is significantly smaller in comparison to the in-plane one \cite{Mob2D}, these structures can be treated as quasi-two-dimensional. The lattice parameters relevant for the further calculations are given in Table \ref{tab:crystals}.  

\begin{table}
\caption{\label{tab:crystals} Lattice constants and angles 
for the unit cells of studied structures}
\begin{tabular}{p{2cm}p{1cm}p{1cm}p{1cm}p{1cm}}
  \hline
  \hline
& a[$\mathring{\rm{A}}$] & b[$\mathring{\rm{A}}$] & $\gamma$ [$^\circ$] & Ref. \rule{0pt}{0.35cm} \\
\hline
pentacene &  6.27 &  7.78 & 87.8 & \cite{Ugao} \\
tetracene  & 6.06 &  7.84 & 85.8 & \cite{Ugaotetracen} \\
picene  & 8.48 &  6.15 & 90 & \cite{Ugaopicen},\cite{GatuzoJPC}  \\
chrysene  & 8.39 & 6.20 & 90 & \cite{Ugaokrizen} \\
\hline
\hline
\end{tabular}
\end{table}

\begin{figure} 
\includegraphics[scale = 0.35]{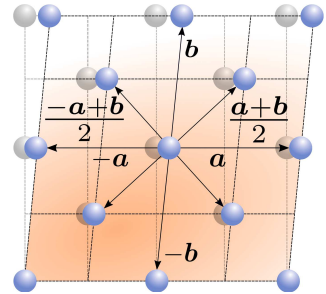}
{\caption{\label{strukture} Schematic presentation of the
analyzed crystal structures: pentacene and tetracene (sketch in color) vs. picene and chrysene (gray-scale sketch). To the each set of lattice vectors  $\{\bm{a},-\bm{a}\}$, $\{\bm{b},-\bm{b}\}$ and $\{\frac{\bm{a}+\bm{b}}{2},
\frac{-\bm{a}+\bm{b}}{2},-\frac{\bm{a}+\bm{b}}{2},-\frac{-\bm{a}+\bm{b}}{2}\}$  corresponds
a pair of exchange integrals (see text).}}
\end{figure}

A 2D sketch of all analyzed crystal structures is shown in Fig 1. In case of phenacenes angle between directions $\bm a$ and $\bm b$ is $90^{\circ}$ (gray-scale sketch), whereas for acenes (colored image) direction $\bm b$ is slightly shifted, i.e. close to $90^{\circ}$ (see Table \ref{tab:crystals}). Therefore, we approximate acenes' lattice by introducing the additional constraint $\bm a \cdot \bm b = 0$ \cite{breakdown,Materials}.
For all structures shown in Fig. \ref{strukture}. central motive has three types of neighbors: two neighbors at points $\bm \lambda_1=\{\bm{a},-\bm{a}\}$ 
coupled trough exchange integral $I_1$, two neighbors at points $\bm \lambda_2 = 
\{\bm{b},-\bm{b}\}$  coupled through exchange integral $I_2$ and four neighbors 
at points $\bm \lambda_3 = \{\frac{\bm{a}+\bm{b}}{2},\frac{-\bm{a}+\bm{b}}{2},
-\frac{\bm{a}+\bm{b}}{2},-\frac{-\bm{a}+\bm{b}}{2}\}$ coupled via exchange integral $I_3$. 
As we have already stated, the mapping between paulion and Heisenberg Hamiltonian
demands anisotropic exchange interactions. As a consequence, all exchange
integrals possess $x$ and $z$ components: $I_{\alpha} \rightarrow (I_{\alpha}^x,I_{\alpha}^z) $,
where $\alpha = 1,2$ or $3$.

\bc
\ec

We have shown that bosonization of the Hamiltonian (\ref{spinham}) in Bloch approximation is sufficient to reproduce exciton dispersion in pentacene in Brillouin zone directions where Davydov splitting is not observed \cite{Materials} . However, in order to reproduce both Davydov components, at each lattice site $\bm{n}$ we define a set of boson occupation states $\{|N_{\rm{A}}\rangle_{\bm{n}}\otimes|N_{\rm{B}}\rangle_{\bm{n}}\}$, 
where indices $\rm{A}$ and $\rm{B}$ refer to different Davydov components.
Therefore, we upgrade Bloch Hamiltonian by introducing direct sum 
\be
\tilde{H}=\tilde{H}_{\rm A}\oplus\tilde{H}_{\rm B}.\label{Hamfin}
\ee
The corresponding Hamiltonians are then defined by 
\be
 \tilde{H}_{\rm{A}}=\tilde{H}'_0+\sum_{\bm{k}}E_{\rm{A}}(\bm{k})B^{\dagger}_{\rm{A}\bm{k}}B_{\rm{A}\bm{k}}\,, 
\label{BlochHam1}
\ee

\be
 \tilde{H}_{\rm{B}}=\tilde{H}'_0+\sum_{\bm{k}}E_{\rm{B}}(\bm{k})B^{\dagger}_{\rm{B}\bm{k}}B_{\rm{B}\bm{k}}\,,
\label{BlochHam2}
\ee

where boson commutation relations read

\bea
 \left[B_{i\bm{k}},B^{\dagger}_{j\bm{q}}\right]&=&\delta_{i,j}\delta_{\bm k,\bm q}\,,\nonumber\\
 \left[B_{i\bm{k}},B_{j\bm{q}}\right]&=&\left[B^{\dagger}_{i\bm{k}},B^{\dagger}_{j\bm{q}}\right]=0\,,\,\,\, i,j=\rm A,\rm B\,.
\label{commutation}
\eea

Exciton dispersion $E_{\rm{A}/{\rm{B}}}(\bm{k})$ is given by 

\bea E_{\rm{{A}}/{\rm{B}}}(\bm{k})&=&\Delta_{\rm{{A}}/{\rm{B}}}-I^x_{1_{\rm{A}/{\rm{B}}}} \cos(\bm{k}\cdot \bm{a})-I_{2_{\rm{A}/{\rm{B}}}}^x \cos(\bm{k}\cdot \bm{b})\nonumber\\
&-&2I_{3_{\rm{A}/{\rm{B}}}}^x \cos\left(\frac{\bm{k} \cdot \bm{a}}{2}\right)\cos\left(\frac{\bm{k} \cdot \bm{b}}{2}\right), \label{Blochdis} \eea
where 
\be
\Delta_{\rm{A}}=\Delta_{\rm{B}}=\Delta=I_1^z+I_2^z+2I_3^z+\mu \mathcal{H}. \label{DeltaGap}
\ee
Since we have shown earlier \cite{Materials} that the influence of the exciton-exciton interaction is negligible in whole temperature range, we can use the same exchange integrals to reproduce experimental data obtained at different temperatures, whereas the gap $\Delta$ changes with temperature due to the change of the parameter $\mathcal{H}$. Experiments show that in Brillouin zone directions along which Davydov splitting is observed the upper Davydov component presents the mirror-like image of the lower one \cite{rothpentacensvi,rothtetracen}. Therefore, we impose that the exchange integrals which correspond to different Davydov components are related by $I_{\alpha_{\rm{B}}}^x=-I_{\alpha_{\rm{A}}}^x,\,\,\alpha=1,2,3$. In next Section we shall present our results for exciton dispersion in pentacene and tetracene (IIIA.) and picene and chrysene (IIIB.).
\bc
\begin{figure}[h] 
\includegraphics[scale=0.6]{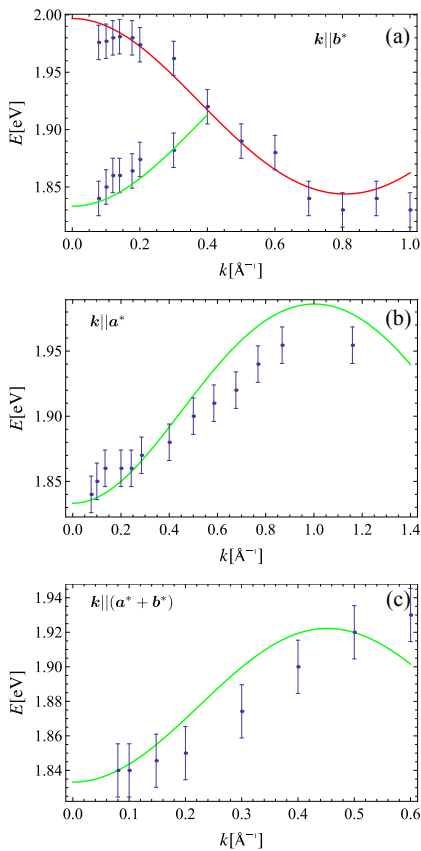} 
{\caption{\label{pentacen20K}
Exciton dispersion in pentacene along three different directions in reciprocal lattice at $T=20\,\rm{K}$. Experimental data are taken 
from \cite{rothpentacensvi}. Theoretical curves are obtained for: $\Delta=1.915\,\rm{eV}$, $I_{1_{\rm{A}}}^x=3.2\,\rm{meV}$, $I_{2_{\rm{A}}}^x=2.2\,\rm{meV}$, $I_{3_{\rm{A}}}^x=38.2\,\rm{meV}$.}}
\end{figure}
\ec
\bc
\begin{figure}[h] 
\includegraphics[scale=0.6]{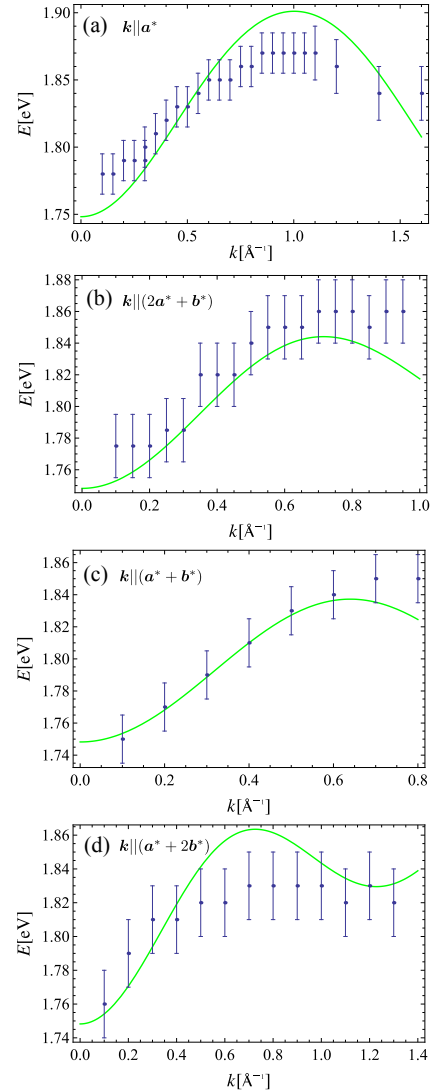} 
{\caption{\label{pentacen300K}
Exciton dispersion in pentacene along four different directions in reciprocal lattice at $T=300\,\rm{K}$. Experimental data are taken 
from \cite{breakdown}. Theoretical curves are obtained for the exchange integral set from Fig. \ref{pentacen20K} and the gap value $\Delta=1.83\,\rm{eV}$.}}
\end{figure}
\ec

\section{Results and discussion}

\subsection{Pentacene and tetracene}

By fitting (\ref{Blochdis}) to experimental data for pentacene taken from \cite{rothpentacensvi} we obtain the following set of parameters: $\Delta=1.915\,\rm{eV}$, $I_{1_{\rm{A}}}^x=3.2\,\rm{meV}$, $I_{2_{\rm{A}}}^x=2.2\,\rm{meV}$, $I_{3_{\rm{A}}}^x=38.2\,\rm{meV}$. Exciton dispersion for this parameter set along different reciprocal $\bm{a}^*\bm{b}^*$ plane directions is shown in Fig. \ref{pentacen20K}. together with the experimental data from \cite{rothpentacensvi}. 
As can be seen from Fig. \ref{pentacen20K}, exciton dispersion obtained without taking CT excitons into account is in a good agreement with the experimental data. Let us note that in Fig. 2(a), where Davydov splitting is observed, upper branch (B) is obtained by inverting the sign of the exchange integrals $x$-components ($I_{\alpha_{\rm{B}}}^x=-I_{\alpha_{\rm{A}}}^x,\,\,\alpha=1,2,3$) of the lower one (A). Hereafter, we shall use green colour for the lower branches and red for the upper ones. With the same exchange integrals we obtain dispersion at room temperature ($T=300\,\rm K$), which is shown in Fig. \ref{pentacen300K} together with the experimental data from \cite{breakdown}. Let us emphasize that since we have used a single set of model
parameters, the plotted dispersion law displays the unique limit, which at $T=20\,\rm{K}$ equals
$\Delta - I_1^x - I_2^x -2 I_3^x = 1.83881 \text{eV}$ as $|\bm k| \to 0$, while at  $T=300\,\rm{K}$ it amounts to $1.75381 \text{eV}$ due to the difference in the gap value $\Delta$.

Theoretical curves in Fig. \ref{pentacen20K},\ref{pentacen300K} will look like straight lines if we plot $E (\bm k)$ in wider energy range. This is due to the fact that $I_{\alpha_{\rm{A/B}}}^x/\Delta \to 0 $. Minor deviations of the theoretical curves from the experimental data are attributed to the presence of other excitations in the system. However, their influence is small in comparison with the Frenkel excitons.

\begin{figure} 
\includegraphics[scale=0.6]{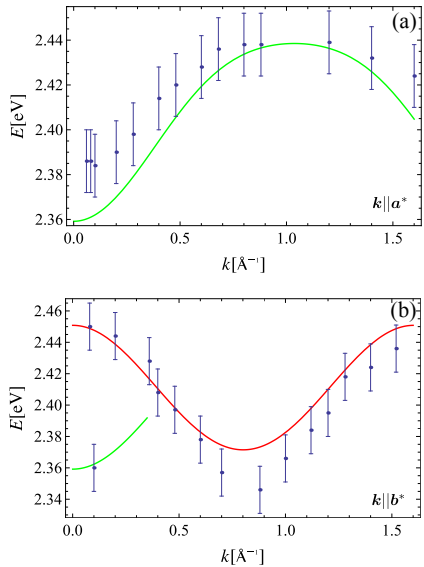} 
{\caption{\label{tetracen}
Exciton dispersion in tetracene along two different directions in reciprocal lattice. Experimental data at $T=20\,\rm{K}$ are taken 
from \cite{rothtetracen}. Theoretical curves are obtained for: $\Delta=2.405\,\rm{eV}$, $I_{1_{\rm{A}}}^x=5.7\,\rm{meV}$, $I_{2_{\rm{A}}}^x=0.4\,\rm{meV}$, $I_{3_{\rm{A}}}^x=19.8\,\rm{meV}$.}}
\end{figure}

Following analogous procedure, we investigate tetracene, another acenes family hydrocarbon with the larger band gap. Using available experimental data from \cite{rothtetracen}, we obtain the corresponding parameter set. Due to the similarity between pentacene and tetracene structures, we use the same exchange integral component inversion rule to reproduce both Davydov branches. Our results along two different Brillouin zone directions together with the available experimental data are shown in Fig. \ref{tetracen}.
\bc
\ec
\par As in pentacene case, theoretical curves periodicity follows experimental data. In the vicinity of $k=0.8\,\mathring{\rm A}^{-1}$ the discrepancy between experimental values and theoretical predictions may originate from the low accuracy with which the measurement was performed in that region \cite{rothtetracen}.
Analysis of Fig. \ref{tetracen} shows that the total band width is roughly twice smaller than in pentacene, leading to smaller variation in the exciton dispersion. Comparing the exchange integral set in pentacene and tetracene, we observe that $|I_{3_{\rm{A/B}}}^x|$ for tetracene is approximately $50\%$ of the same parameter value for pentacene. It should be noted that $|I_{3_{\rm{A/B}}}^x|$ is directly proportional to the exciton mobility \cite{Materials}. With the increase of the optical gap $\Delta$, the mobility decreases and so does the parameter $|I_{3_{\rm{A/B}}}^x|$. Therefore, the value of this parameter for different aromatic hydrocarbons can be used as a consistency check of our calculations. In order to make our conclusions concerning the magnitudes of the optical (and band) gaps and the corresponding $|I_{3}^x|$ values in analyzed hydrocarbons more transparent, we compare them in Table \ref{tab:gepovi}.

\begin{table}
\caption{\label{tab:gepovi} Transport energy gaps ($E_{\rm g}$) for studied structures vs. calculated optical gaps ($\Delta$) together with the corresponding $|I_{3}^x|$ values (at $T=20\,\rm{K}$)}
\begin{tabular}{p{2cm}p{2cm}p{1.5cm}p{1.5cm}}
  \hline
  \hline
& $E_{\rm g}\, [\rm{eV}]$ & $\Delta\, [\rm{eV}]$ & $|I_{3}^x|\,[\rm{meV}]$ \\
\hline
pentacene &  2.2 \cite{amy,sato2} &  1.915 & 38.2  \\
tetracene  & 3.3 \cite{amy,sato2} &  2.405 & 19.8  \\
picene  & 4.05 \cite{picengep,sato} &  3.249 & 2.8   \\
chrysene  & 4.2 \cite{sato} & 3.4 & 2.8  \\
\hline
\hline
\end{tabular}
\end{table}

\par
\begin{figure} 
\includegraphics[scale=0.6]{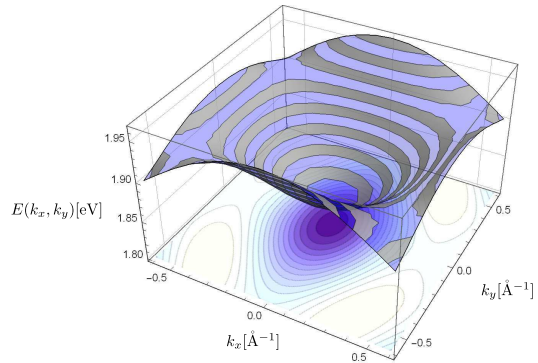} 
{\caption{\label{pentacen3D}
3D plot of exciton dispersion in pentacene at $T=20\,\rm{K}$. Parameter set is the same as in Fig. \ref{pentacen20K}.}}
\end{figure}

 Finally, it is possible to draw a three-dimensional plot of the exciton dispersion in acenes. As an example,
exciton dispersion $E(k_x,k_y)$ in pentacene at $T=20\,\rm K$ is given in Fig. \ref{pentacen3D}.
\bc
\ec

\bc
\begin{figure} 
\includegraphics[scale=0.6]{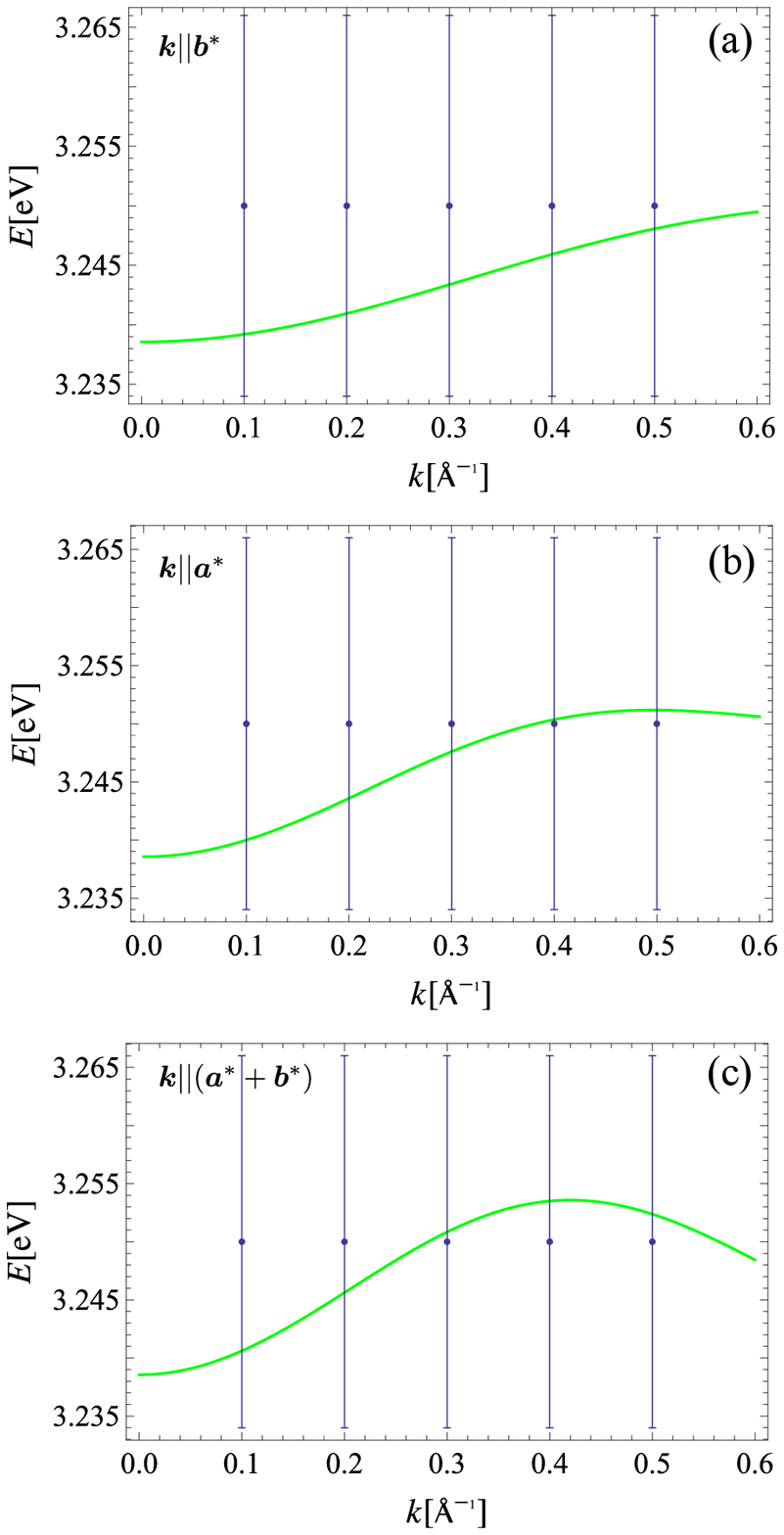}
{\caption{\label{picen}
Exciton dispersion in picene along three different directions in reciprocal lattice. Experimental data at $T=20\,\rm K$ are taken 
from \cite{rothkrizen}. Theoretical curves are obtained for: $\Delta=3.249\,\rm{eV}$, $I_{1_{\rm{A}}}^x=2.8\,\rm{meV}$, $I_{2_{\rm{A}}}^x=2\,\rm{meV}$, $I_{3_{\rm{A}}}^x=2.8\,\rm{meV}$.}}
\end{figure}
\ec

\bc
\begin{figure} 
\includegraphics[scale=0.6]{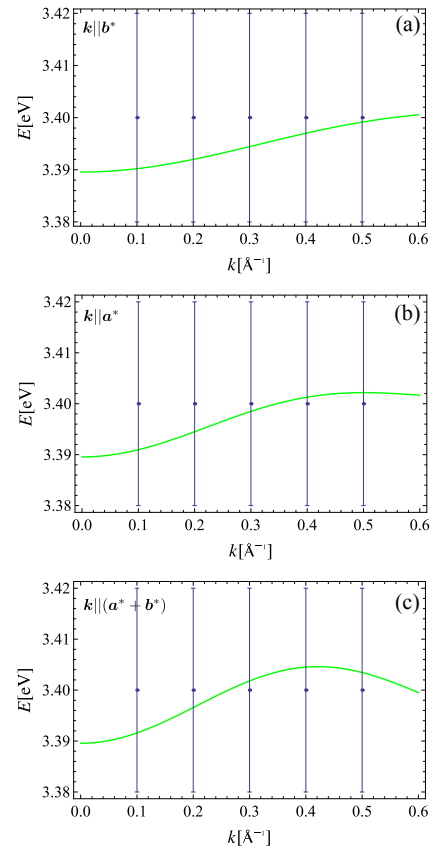}
{\caption{\label{krizen}
Exciton dispersion in chrysene along three different directions in reciprocal lattice. Experimental data at $T=20\,\rm K$ are taken 
from \cite{rothkrizen}. Theoretical curves are obtained for: $\Delta=3.4\,\rm{eV}$, $I_{1_{\rm{A}}}^x=2.8\,\rm{meV}$, $I_{2_{\rm{A}}}^x=2\,\rm{meV}$, $I_{3_{\rm{A}}}^x=2.8\,\rm{meV}$.}}
\end{figure}
\ec

\subsection{Picene and chrysene}

Applying the same procedure for the phenacenes picene and chrysene, we obtain the exciton dispersion within the reciprocal $\bm{a}^*\bm{b}^*$ plane and compare it to the experimental data from \cite{rothkrizen,Ugaopicen} (see Figs. \ref{picen},\ref{krizen}).  Due to the pronounced similarity between the crystal structures of these molecular solids, we were able to reproduce the experimental data for both hydrocarbons in satisfactory manner with the same exchange integrals set. Comparing the exchange integrals $x$-components with those for acenes, we notice that in case of phenacenes the value of $I_{3_{\rm{A}}}^x$ is significantly smaller. This is attributed to the fact that the band gap (and consequently the optical gap) in phenacenes is larger than in acenes (see Table \ref{tab:gepovi}). 
\par By inspection of Figs. \ref{picen},\ref{krizen} we notice that dispersion in case of phenacenes is more isotropic. Further, the total band width is approximately $10\,\rm{meV}$, which is roughly ten times smaller value than for acenes. 

\begin{figure} 
\includegraphics[scale=0.6]{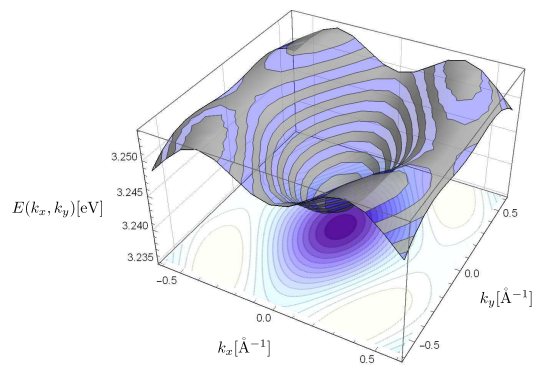} 
{\caption{\label{picen3D}
3D plot of exciton dispersion in picene, obtained with the parameters from Fig. \ref{picen}.}}
\end{figure}

Analogous to IIIA. we present the three-dimensional plot of exciton dispersion in picene (Fig. \ref{picen3D}).
\bc
\ec

\section{Conclusion} 
We analyze the exciton dispersion in different molecular solids relying on the correspondence between Pauli (\ref{paulham}) and
Heisenberg (\ref{spinham}) Hamiltonians. Following standard Bloch bosonization procedure we obtain exciton dispersion relation. In order to investigate Davydov splitting phenomenon, we use the upgraded model Hamiltonian introduced in this paper (\ref{Hamfin}). By fitting exchange integrals
to the experimental results, we obtain exciton dispersions that possess the same periodicity as experimental data.
Analyses of dispersion curves show that in the acenes (pentacene and tetracene) the dispersion is rather anisotropic, contrary to the phenacenes (picene and chrysene) where it is more isotropic and almost constant. As regards the phenacenes, our results corroborate the earlier stated conclusion \cite{Ugaopicen,rothkrizen} that the lowest-lying excitations in picene and chrysene are localized Frenkel excitons. However, we obtain that the experimental data for pentacene and tetracene can also be satisfactorily reproduced within the noninteracting exciton picture. Therefore, we suggest that the influence of the CT excitons on exciton dispersion in acenes is not as large as claimed earlier.
Further, it can be seen that the dispersion curves along different Brillouin zone directions for given molecular solid tend to the same value as $k\to 0$. This reflects the fact that the single parameter set is used for all $\bm k$-space directions, unlike \cite{breakdown}.  
Comparative study of analyzed hydrocarbons shows that the magnitudes of the optical gap $\Delta$ and the exchange integral $|I_3^x|$ are inversely proportional.  
Therefore, in phenacenes the carrier mobility is smaller, i.e. the exciton-exciton interaction is weaker than in acenes. However, despite those differences, our calculations based on Frenkel exciton model reproduce the experimental data for all analyzed hydrocarbons in satisfying manner. This suggests that small variations of the exciton dispersion should not be connected with the applicability of the Frenkel exciton model. In order to corroborate this statement, a variety of examples can be found in the magnon dispersion analyses for different magnetic insulators \cite{primerPRL,primerEuO,primer2EuO,primerSloba}.
In order to further test our suggested model, additional measurements of exciton dispersion where Davydov splitting is observed are required.

\section*{Acknowledgments}
This work was supported by the Serbian Ministry of
Education and Science under Contract No. OI-171009
and by the Provincial Secretariat for High Education and Scientific 
Research of Vojvodina (Project No. APV 114-451-2201).

\bibliography{Refs}

\begin{thebibliography}{45}%
\makeatletter
\providecommand \@ifxundefined [1]{%
 \@ifx{#1\undefined}
}%
\providecommand \@ifnum [1]{%
 \ifnum #1\expandafter \@firstoftwo
 \else \expandafter \@secondoftwo
 \fi
}%
\providecommand \@ifx [1]{%
 \ifx #1\expandafter \@firstoftwo
 \else \expandafter \@secondoftwo
 \fi
}%
\providecommand \natexlab [1]{#1}%
\providecommand \enquote  [1]{``#1''}%
\providecommand \bibnamefont  [1]{#1}%
\providecommand \bibfnamefont [1]{#1}%
\providecommand \citenamefont [1]{#1}%
\providecommand \href@noop [0]{\@secondoftwo}%
\providecommand \href [0]{\begingroup \@sanitize@url \@href}%
\providecommand \@href[1]{\@@startlink{#1}\@@href}%
\providecommand \@@href[1]{\endgroup#1\@@endlink}%
\providecommand \@sanitize@url [0]{\catcode `\\12\catcode `\$12\catcode
  `\&12\catcode `\#12\catcode `\^12\catcode `\_12\catcode `\%12\relax}%
\providecommand \@@startlink[1]{}%
\providecommand \@@endlink[0]{}%
\providecommand \url  [0]{\begingroup\@sanitize@url \@url }%
\providecommand \@url [1]{\endgroup\@href {#1}{\urlprefix }}%
\providecommand \urlprefix  [0]{URL }%
\providecommand \Eprint [0]{\href }%
\providecommand \doibase [0]{http://dx.doi.org/}%
\providecommand \selectlanguage [0]{\@gobble}%
\providecommand \bibinfo  [0]{\@secondoftwo}%
\providecommand \bibfield  [0]{\@secondoftwo}%
\providecommand \translation [1]{[#1]}%
\providecommand \BibitemOpen [0]{}%
\providecommand \bibitemStop [0]{}%
\providecommand \bibitemNoStop [0]{.\EOS\space}%
\providecommand \EOS [0]{\spacefactor3000\relax}%
\providecommand \BibitemShut  [1]{\csname bibitem#1\endcsname}%
\let\auto@bib@innerbib\@empty
\bibitem [{\citenamefont {Gershenson}\ \emph {et~al.}(2006)\citenamefont
  {Gershenson}, \citenamefont {Podzorov},\ and\ \citenamefont
  {Morpurgo}}]{tranzistor}%
  \BibitemOpen
  \bibfield  {author} {\bibinfo {author} {\bibfnamefont {M.}~\bibnamefont
  {Gershenson}}, \bibinfo {author} {\bibfnamefont {V.}~\bibnamefont
  {Podzorov}}, \ and\ \bibinfo {author} {\bibfnamefont {A.}~\bibnamefont
  {Morpurgo}},\ }\href
  {https://journals.aps.org/rmp/abstract/10.1103/RevModPhys.78.973} {\bibfield
  {journal} {\bibinfo  {journal} {Rev. Mod. Phys.}\ }\textbf {\bibinfo {volume}
  {\textbf{78}}},\ \bibinfo {pages} {973} (\bibinfo {year} {2006})}\BibitemShut
  {NoStop}%
\bibitem [{\citenamefont {Peumans}\ \emph {et~al.}(2011)\citenamefont
  {Peumans}, \citenamefont {Uchida},\ and\ \citenamefont
  {Forrest}}]{naturePeumans}%
  \BibitemOpen
  \bibfield  {author} {\bibinfo {author} {\bibfnamefont {P.}~\bibnamefont
  {Peumans}}, \bibinfo {author} {\bibfnamefont {S.}~\bibnamefont {Uchida}}, \
  and\ \bibinfo {author} {\bibfnamefont {S.~R.}\ \bibnamefont {Forrest}},\ }in\
  \href {https://www.worldscientific.com/doi/abs/10.1142/9789814317665_0015}
  {\emph {\bibinfo {booktitle} {Materials for Sustainable Energy: A Collection
  of Peer-Reviewed Research and Review Articles from Nature Publishing
  Group}}}\ (\bibinfo  {publisher} {World Scientific},\ \bibinfo {year}
  {2011})\ pp.\ \bibinfo {pages} {94--98}\BibitemShut {NoStop}%
\bibitem [{\citenamefont {Baldo}\ \emph {et~al.}(1998)\citenamefont {Baldo},
  \citenamefont {O'brien}, \citenamefont {You}, \citenamefont {Shoustikov},
  \citenamefont {Sibley}, \citenamefont {Thompson},\ and\ \citenamefont
  {Forrest}}]{natureBaldo}%
  \BibitemOpen
  \bibfield  {author} {\bibinfo {author} {\bibfnamefont {M.~A.}\ \bibnamefont
  {Baldo}}, \bibinfo {author} {\bibfnamefont {D.}~\bibnamefont {O'brien}},
  \bibinfo {author} {\bibfnamefont {Y.}~\bibnamefont {You}}, \bibinfo {author}
  {\bibfnamefont {A.}~\bibnamefont {Shoustikov}}, \bibinfo {author}
  {\bibfnamefont {S.}~\bibnamefont {Sibley}}, \bibinfo {author} {\bibfnamefont
  {M.}~\bibnamefont {Thompson}}, \ and\ \bibinfo {author} {\bibfnamefont
  {S.~R.}\ \bibnamefont {Forrest}},\ }\href
  {https://www.nature.com/articles/25954} {\bibfield  {journal} {\bibinfo
  {journal} {Nature}\ }\textbf {\bibinfo {volume} {395}},\ \bibinfo {pages}
  {151} (\bibinfo {year} {1998})}\BibitemShut {NoStop}%
\bibitem [{\citenamefont {Forrest}(2004)}]{forest}%
  \BibitemOpen
  \bibfield  {author} {\bibinfo {author} {\bibfnamefont {S.~R.}\ \bibnamefont
  {Forrest}},\ }\href {https://www.nature.com/articles/nature02498} {\bibfield
  {journal} {\bibinfo  {journal} {Nature}\ }\textbf {\bibinfo {volume} {428}},\
  \bibinfo {pages} {911} (\bibinfo {year} {2004})}\BibitemShut {NoStop}%
\bibitem [{\citenamefont {Li}\ \emph {et~al.}(2011)\citenamefont {Li},
  \citenamefont {Shrotriya}, \citenamefont {Huang}, \citenamefont {Yao},
  \citenamefont {Moriarty}, \citenamefont {Emery},\ and\ \citenamefont
  {Yang}}]{jang}%
  \BibitemOpen
  \bibfield  {author} {\bibinfo {author} {\bibfnamefont {G.}~\bibnamefont
  {Li}}, \bibinfo {author} {\bibfnamefont {V.}~\bibnamefont {Shrotriya}},
  \bibinfo {author} {\bibfnamefont {J.}~\bibnamefont {Huang}}, \bibinfo
  {author} {\bibfnamefont {Y.}~\bibnamefont {Yao}}, \bibinfo {author}
  {\bibfnamefont {T.}~\bibnamefont {Moriarty}}, \bibinfo {author}
  {\bibfnamefont {K.}~\bibnamefont {Emery}}, \ and\ \bibinfo {author}
  {\bibfnamefont {Y.}~\bibnamefont {Yang}},\ }\href
  {https://www.nature.com/articles/nmat1500} {\bibfield  {journal} {\bibinfo
  {journal} {Nat. Mater.}\ }\textbf {\bibinfo {volume} {4}},\ \bibinfo {pages}
  {864} (\bibinfo {year} {2011})}\BibitemShut {NoStop}%
\bibitem [{\citenamefont {Mitsuhashi}\ \emph {et~al.}(2010)\citenamefont
  {Mitsuhashi}, \citenamefont {Suzuki}, \citenamefont {Yamanari}, \citenamefont
  {Mitamura}, \citenamefont {Kambe}, \citenamefont {Ikeda}, \citenamefont
  {Okamoto}, \citenamefont {Fujiwara}, \citenamefont {Yamaji}, \citenamefont
  {Kawasaki} \emph {et~al.}}]{nature}%
  \BibitemOpen
  \bibfield  {author} {\bibinfo {author} {\bibfnamefont {R.}~\bibnamefont
  {Mitsuhashi}}, \bibinfo {author} {\bibfnamefont {Y.}~\bibnamefont {Suzuki}},
  \bibinfo {author} {\bibfnamefont {Y.}~\bibnamefont {Yamanari}}, \bibinfo
  {author} {\bibfnamefont {H.}~\bibnamefont {Mitamura}}, \bibinfo {author}
  {\bibfnamefont {T.}~\bibnamefont {Kambe}}, \bibinfo {author} {\bibfnamefont
  {N.}~\bibnamefont {Ikeda}}, \bibinfo {author} {\bibfnamefont
  {H.}~\bibnamefont {Okamoto}}, \bibinfo {author} {\bibfnamefont
  {A.}~\bibnamefont {Fujiwara}}, \bibinfo {author} {\bibfnamefont
  {M.}~\bibnamefont {Yamaji}}, \bibinfo {author} {\bibfnamefont
  {N.}~\bibnamefont {Kawasaki}},  \emph {et~al.},\ }\href
  {https://www.nature.com/articles/nature08859} {\bibfield  {journal} {\bibinfo
   {journal} {Nature}\ }\textbf {\bibinfo {volume} {464}},\ \bibinfo {pages}
  {76} (\bibinfo {year} {2010})}\BibitemShut {NoStop}%
\bibitem [{\citenamefont {Knupfer}\ and\ \citenamefont {Berger}(2006)}]{183}%
  \BibitemOpen
  \bibfield  {author} {\bibinfo {author} {\bibfnamefont {M.}~\bibnamefont
  {Knupfer}}\ and\ \bibinfo {author} {\bibfnamefont {H.}~\bibnamefont
  {Berger}},\ }\href
  {http://www.sciencedirect.com/science/article/pii/S0301010405002971}
  {\bibfield  {journal} {\bibinfo  {journal} {Chem. Phys.}\ }\textbf {\bibinfo
  {volume} {325}},\ \bibinfo {pages} {92 } (\bibinfo {year}
  {2006})}\BibitemShut {NoStop}%
\bibitem [{\citenamefont {Schuster}\ \emph {et~al.}(2007)\citenamefont
  {Schuster}, \citenamefont {Knupfer},\ and\ \citenamefont
  {Berger}}]{breakdown}%
  \BibitemOpen
  \bibfield  {author} {\bibinfo {author} {\bibfnamefont {R.}~\bibnamefont
  {Schuster}}, \bibinfo {author} {\bibfnamefont {M.}~\bibnamefont {Knupfer}}, \
  and\ \bibinfo {author} {\bibfnamefont {H.}~\bibnamefont {Berger}},\ }\href
  {\doibase 10.1103/PhysRevLett.98.037402} {\bibfield  {journal} {\bibinfo
  {journal} {Phys. Rev. Lett.}\ }\textbf {\bibinfo {volume} {98}},\ \bibinfo
  {pages} {037402} (\bibinfo {year} {2007})}\BibitemShut {NoStop}%
\bibitem [{\citenamefont {Roth}\ \emph {et~al.}(2011)\citenamefont {Roth},
  \citenamefont {Mahns}, \citenamefont {B\"uchner},\ and\ \citenamefont
  {Knupfer}}]{Ugaopicen}%
  \BibitemOpen
  \bibfield  {author} {\bibinfo {author} {\bibfnamefont {F.}~\bibnamefont
  {Roth}}, \bibinfo {author} {\bibfnamefont {B.}~\bibnamefont {Mahns}},
  \bibinfo {author} {\bibfnamefont {B.}~\bibnamefont {B\"uchner}}, \ and\
  \bibinfo {author} {\bibfnamefont {M.}~\bibnamefont {Knupfer}},\ }\href
  {https://journals.aps.org/prb/abstract/10.1103/PhysRevB.83.165436} {\bibfield
   {journal} {\bibinfo  {journal} {{Phys. Rev. B}}\ }\textbf {\bibinfo {volume}
  {\text{83}}},\ \bibinfo {pages} {165436} (\bibinfo {year}
  {2011})}\BibitemShut {NoStop}%
\bibitem [{\citenamefont {Roth}\ \emph
  {et~al.}(2012{\natexlab{a}})\citenamefont {Roth}, \citenamefont {Schuster},
  \citenamefont {K{\"o}nig}, \citenamefont {Knupfer},\ and\ \citenamefont
  {Berger}}]{rothpentacensvi}%
  \BibitemOpen
  \bibfield  {author} {\bibinfo {author} {\bibfnamefont {F.}~\bibnamefont
  {Roth}}, \bibinfo {author} {\bibfnamefont {R.}~\bibnamefont {Schuster}},
  \bibinfo {author} {\bibfnamefont {A.}~\bibnamefont {K{\"o}nig}}, \bibinfo
  {author} {\bibfnamefont {M.}~\bibnamefont {Knupfer}}, \ and\ \bibinfo
  {author} {\bibfnamefont {H.}~\bibnamefont {Berger}},\ }\href
  {https://aip.scitation.org/doi/10.1063/1.4723812} {\bibfield  {journal}
  {\bibinfo  {journal} {J. Chem. Phys.}\ }\textbf {\bibinfo {volume} {136}},\
  \bibinfo {pages} {204708} (\bibinfo {year} {2012}{\natexlab{a}})}\BibitemShut
  {NoStop}%
\bibitem [{\citenamefont {Roth}\ \emph {et~al.}(2013)\citenamefont {Roth},
  \citenamefont {Mahns}, \citenamefont {Hampel}, \citenamefont {Nohr},
  \citenamefont {Berger}, \citenamefont {B{\"u}chner},\ and\ \citenamefont
  {Knupfer}}]{rothlepaslika}%
  \BibitemOpen
  \bibfield  {author} {\bibinfo {author} {\bibfnamefont {F.}~\bibnamefont
  {Roth}}, \bibinfo {author} {\bibfnamefont {B.}~\bibnamefont {Mahns}},
  \bibinfo {author} {\bibfnamefont {S.}~\bibnamefont {Hampel}}, \bibinfo
  {author} {\bibfnamefont {M.}~\bibnamefont {Nohr}}, \bibinfo {author}
  {\bibfnamefont {H.}~\bibnamefont {Berger}}, \bibinfo {author} {\bibfnamefont
  {B.}~\bibnamefont {B{\"u}chner}}, \ and\ \bibinfo {author} {\bibfnamefont
  {M.}~\bibnamefont {Knupfer}},\ }\href
  {https://link.springer.com/article/10.1140/epjb/e2012-30592-1} {\bibfield
  {journal} {\bibinfo  {journal} {Eur. Phys. J. B}\ }\textbf {\bibinfo {volume}
  {86}},\ \bibinfo {pages} {66} (\bibinfo {year} {2013})}\BibitemShut {NoStop}%
\bibitem [{\citenamefont {Roth}\ \emph {et~al.}(2015)\citenamefont {Roth},
  \citenamefont {Nohr}, \citenamefont {Hampel},\ and\ \citenamefont
  {Knupfer}}]{rothtetracen}%
  \BibitemOpen
  \bibfield  {author} {\bibinfo {author} {\bibfnamefont {F.}~\bibnamefont
  {Roth}}, \bibinfo {author} {\bibfnamefont {M.}~\bibnamefont {Nohr}}, \bibinfo
  {author} {\bibfnamefont {S.}~\bibnamefont {Hampel}}, \ and\ \bibinfo {author}
  {\bibfnamefont {M.}~\bibnamefont {Knupfer}},\ }\href
  {https://iopscience.iop.org/article/10.1209/0295-5075/112/37004} {\bibfield
  {journal} {\bibinfo  {journal} {EPL}\ }\textbf {\bibinfo {volume} {112}},\
  \bibinfo {pages} {37004} (\bibinfo {year} {2015})}\BibitemShut {NoStop}%
\bibitem [{\citenamefont {Roth}\ and\ \citenamefont
  {Knupfer}(2015)}]{rothkrizen}%
  \BibitemOpen
  \bibfield  {author} {\bibinfo {author} {\bibfnamefont {F.}~\bibnamefont
  {Roth}}\ and\ \bibinfo {author} {\bibfnamefont {M.}~\bibnamefont {Knupfer}},\
  }\href {https://www.sciencedirect.com/science/article/pii/S0368204815000225}
  {\bibfield  {journal} {\bibinfo  {journal} {J. Electron Spectrosc.}\ }\textbf
  {\bibinfo {volume} {204}},\ \bibinfo {pages} {23} (\bibinfo {year}
  {2015})}\BibitemShut {NoStop}%
\bibitem [{\citenamefont {Cudazzo}\ \emph {et~al.}(2012)\citenamefont
  {Cudazzo}, \citenamefont {Gatti},\ and\ \citenamefont {Rubio}}]{GatuzoPRB}%
  \BibitemOpen
  \bibfield  {author} {\bibinfo {author} {\bibfnamefont {P.}~\bibnamefont
  {Cudazzo}}, \bibinfo {author} {\bibfnamefont {M.}~\bibnamefont {Gatti}}, \
  and\ \bibinfo {author} {\bibfnamefont {A.}~\bibnamefont {Rubio}},\ }\href
  {\doibase 10.1103/PhysRevB.86.195307} {\bibfield  {journal} {\bibinfo
  {journal} {Phys. Rev. B}\ }\textbf {\bibinfo {volume} {86}},\ \bibinfo
  {pages} {195307} (\bibinfo {year} {2012})}\BibitemShut {NoStop}%
\bibitem [{\citenamefont {Cudazzo}\ \emph {et~al.}(2013)\citenamefont
  {Cudazzo}, \citenamefont {Gatti}, \citenamefont {Rubio},\ and\ \citenamefont
  {Sottile}}]{GatuzoPRB2}%
  \BibitemOpen
  \bibfield  {author} {\bibinfo {author} {\bibfnamefont {P.}~\bibnamefont
  {Cudazzo}}, \bibinfo {author} {\bibfnamefont {M.}~\bibnamefont {Gatti}},
  \bibinfo {author} {\bibfnamefont {A.}~\bibnamefont {Rubio}}, \ and\ \bibinfo
  {author} {\bibfnamefont {F.}~\bibnamefont {Sottile}},\ }\href {\doibase
  10.1103/PhysRevB.88.195152} {\bibfield  {journal} {\bibinfo  {journal} {Phys.
  Rev. B}\ }\textbf {\bibinfo {volume} {88}},\ \bibinfo {pages} {195152}
  (\bibinfo {year} {2013})}\BibitemShut {NoStop}%
\bibitem [{\citenamefont {Cudazzo}\ \emph {et~al.}(2015)\citenamefont
  {Cudazzo}, \citenamefont {Sottile}, \citenamefont {Rubio},\ and\
  \citenamefont {Gatti}}]{GatuzoJPC}%
  \BibitemOpen
  \bibfield  {author} {\bibinfo {author} {\bibfnamefont {P.}~\bibnamefont
  {Cudazzo}}, \bibinfo {author} {\bibfnamefont {F.}~\bibnamefont {Sottile}},
  \bibinfo {author} {\bibfnamefont {A.}~\bibnamefont {Rubio}}, \ and\ \bibinfo
  {author} {\bibfnamefont {M.}~\bibnamefont {Gatti}},\ }\href
  {http://stacks.iop.org/0953-8984/27/i=11/a=113204} {\bibfield  {journal}
  {\bibinfo  {journal} {J. Phys.: Condens. Matter}\ }\textbf {\bibinfo {volume}
  {27}},\ \bibinfo {pages} {113204} (\bibinfo {year} {2015})}\BibitemShut
  {NoStop}%
\bibitem [{\citenamefont {Kronik}\ and\ \citenamefont {Neaton}(2016)}]{AnnRev}%
  \BibitemOpen
  \bibfield  {author} {\bibinfo {author} {\bibfnamefont {L.}~\bibnamefont
  {Kronik}}\ and\ \bibinfo {author} {\bibfnamefont {J.~B.}\ \bibnamefont
  {Neaton}},\ }\href {https://doi.org/10.1146/annurev-physchem-040214-121351}
  {\bibfield  {journal} {\bibinfo  {journal} {Annu. Rev. Phys. Chem.}\ }\textbf
  {\bibinfo {volume} {67}},\ \bibinfo {pages} {587} (\bibinfo {year}
  {2016})}\BibitemShut {NoStop}%
\bibitem [{\citenamefont {Sharifzadeh}\ \emph {et~al.}(2013)\citenamefont
  {Sharifzadeh}, \citenamefont {Darancet}, \citenamefont {Kronik},\ and\
  \citenamefont {Neaton}}]{JPCLett}%
  \BibitemOpen
  \bibfield  {author} {\bibinfo {author} {\bibfnamefont {S.}~\bibnamefont
  {Sharifzadeh}}, \bibinfo {author} {\bibfnamefont {P.}~\bibnamefont
  {Darancet}}, \bibinfo {author} {\bibfnamefont {L.}~\bibnamefont {Kronik}}, \
  and\ \bibinfo {author} {\bibfnamefont {J.~B.}\ \bibnamefont {Neaton}},\
  }\href {https://doi.org/10.1021/jz401069f} {\bibfield  {journal} {\bibinfo
  {journal} {J. Phys. Chem. Lett.}\ }\textbf {\bibinfo {volume} {4}},\ \bibinfo
  {pages} {2197} (\bibinfo {year} {2013})}\BibitemShut {NoStop}%
\bibitem [{\citenamefont {Gombar}\ \emph {et~al.}(2018)\citenamefont {Gombar},
  \citenamefont {Mali}, \citenamefont {Panti{\'c}}, \citenamefont
  {Pavkov-Hrvojevi{\'c}},\ and\ \citenamefont
  {Rado{\v{s}}evi{\'c}}}]{Materials}%
  \BibitemOpen
  \bibfield  {author} {\bibinfo {author} {\bibfnamefont {S.}~\bibnamefont
  {Gombar}}, \bibinfo {author} {\bibfnamefont {P.}~\bibnamefont {Mali}},
  \bibinfo {author} {\bibfnamefont {M.}~\bibnamefont {Panti{\'c}}}, \bibinfo
  {author} {\bibfnamefont {M.}~\bibnamefont {Pavkov-Hrvojevi{\'c}}}, \ and\
  \bibinfo {author} {\bibfnamefont {S.}~\bibnamefont {Rado{\v{s}}evi{\'c}}},\
  }\href {https://www.mdpi.com/1996-1944/11/11/2219} {\bibfield  {journal}
  {\bibinfo  {journal} {Materials}\ }\textbf {\bibinfo {volume} {11}},\
  \bibinfo {pages} {2219} (\bibinfo {year} {2018})}\BibitemShut {NoStop}%
\bibitem [{\citenamefont {Davydov}(1964)}]{davydov}%
  \BibitemOpen
  \bibfield  {author} {\bibinfo {author} {\bibfnamefont {A.~S.}\ \bibnamefont
  {Davydov}},\ }\href
  {https://iopscience.iop.org/article/10.1070/PU1964v007n02ABEH003659/pdf}
  {\bibfield  {journal} {\bibinfo  {journal} {Soviet Physics Uspekhi}\ }\textbf
  {\bibinfo {volume} {7}},\ \bibinfo {pages} {145} (\bibinfo {year}
  {1964})}\BibitemShut {NoStop}%
\bibitem [{\citenamefont {Agranovich}(2008)}]{Agranovicbook}%
  \BibitemOpen
  \bibfield  {author} {\bibinfo {author} {\bibfnamefont {V.}~\bibnamefont
  {Agranovich}},\ }\href
  {https://www.oxfordscholarship.com/view/10.1093/acprof:oso/9780199234417.001.0001/acprof-9780199234417}
  {\emph {\bibinfo {title} {Excitations in organic solids}}}\ (\bibinfo
  {publisher} {Oxford University Press},\ \bibinfo {year} {2008})\BibitemShut
  {NoStop}%
\bibitem [{\citenamefont {Agranovich}\ and\ \citenamefont
  {Toshich}(1968)}]{Tole}%
  \BibitemOpen
  \bibfield  {author} {\bibinfo {author} {\bibfnamefont {V.}~\bibnamefont
  {Agranovich}}\ and\ \bibinfo {author} {\bibfnamefont {B.}~\bibnamefont
  {Toshich}},\ }\href {http://www.jetp.ac.ru/cgi-bin/dn/e_026_01_0104.pdf}
  {\bibfield  {journal} {\bibinfo  {journal} {Sov. Phys. JETP}\ }\textbf
  {\bibinfo {volume} {26}},\ \bibinfo {pages} {104} (\bibinfo {year}
  {1968})}\BibitemShut {NoStop}%
\bibitem [{\citenamefont {Tyablikov}(1967)}]{Tjablikov}%
  \BibitemOpen
  \bibfield  {author} {\bibinfo {author} {\bibfnamefont {S.~V.}\ \bibnamefont
  {Tyablikov}},\ }\href {https://www.springer.com/gp/book/9781489970916} {\emph
  {\bibinfo {title} {Methods in the Quantum Theory of Magnetism}}}\ (\bibinfo
  {publisher} {Springer},\ \bibinfo {year} {1967})\BibitemShut {NoStop}%
\bibitem [{\citenamefont {Fröbrich}\ and\ \citenamefont
  {Kuntz}(2006)}]{Kuntz}%
  \BibitemOpen
  \bibfield  {author} {\bibinfo {author} {\bibfnamefont {P.}~\bibnamefont
  {Fröbrich}}\ and\ \bibinfo {author} {\bibfnamefont {P.}~\bibnamefont
  {Kuntz}},\ }\href {\doibase https://doi.org/10.1016/j.physrep.2006.07.002}
  {\bibfield  {journal} {\bibinfo  {journal} {Phys. Rep.}\ }\textbf {\bibinfo
  {volume} {432}},\ \bibinfo {pages} {223 } (\bibinfo {year}
  {2006})}\BibitemShut {NoStop}%
\bibitem [{\citenamefont {Auerbach}(2012)}]{Averbah}%
  \BibitemOpen
  \bibfield  {author} {\bibinfo {author} {\bibfnamefont {A.}~\bibnamefont
  {Auerbach}},\ }\href {https://www.springer.com/gp/book/9780387942865} {\emph
  {\bibinfo {title} {Interacting electrons and quantum magnetism}}}\ (\bibinfo
  {publisher} {Springer},\ \bibinfo {year} {2012})\BibitemShut {NoStop}%
\bibitem [{\citenamefont {Manousakis}(1991)}]{Mano}%
  \BibitemOpen
  \bibfield  {author} {\bibinfo {author} {\bibfnamefont {E.}~\bibnamefont
  {Manousakis}},\ }\href {\doibase 10.1103/RevModPhys.63.1} {\bibfield
  {journal} {\bibinfo  {journal} {Rev. Mod. Phys.}\ }\textbf {\bibinfo {volume}
  {63}},\ \bibinfo {pages} {1} (\bibinfo {year} {1991})}\BibitemShut {NoStop}%
\bibitem [{\citenamefont {Nolting}\ and\ \citenamefont
  {Ramakanth}(2009)}]{Nolting}%
  \BibitemOpen
  \bibfield  {author} {\bibinfo {author} {\bibfnamefont {W.}~\bibnamefont
  {Nolting}}\ and\ \bibinfo {author} {\bibfnamefont {A.}~\bibnamefont
  {Ramakanth}},\ }\href {https://www.springer.com/gp/book/9783540854159} {\emph
  {\bibinfo {title} {Quantum theory of magnetism}}}\ (\bibinfo  {publisher}
  {Springer},\ \bibinfo {year} {2009})\BibitemShut {NoStop}%
\bibitem [{\citenamefont {Sandvik}\ and\ \citenamefont
  {Kurkij\"arvi}(1991)}]{Sandvik}%
  \BibitemOpen
  \bibfield  {author} {\bibinfo {author} {\bibfnamefont {A.~W.}\ \bibnamefont
  {Sandvik}}\ and\ \bibinfo {author} {\bibfnamefont {J.}~\bibnamefont
  {Kurkij\"arvi}},\ }\href {\doibase 10.1103/PhysRevB.43.5950} {\bibfield
  {journal} {\bibinfo  {journal} {Phys. Rev. B}\ }\textbf {\bibinfo {volume}
  {43}},\ \bibinfo {pages} {5950} (\bibinfo {year} {1991})}\BibitemShut
  {NoStop}%
\bibitem [{\citenamefont {Panti\'{c}}\ \emph {et~al.}(2014)\citenamefont
  {Panti\'{c}}, \citenamefont {Kapor}, \citenamefont {Rado\v{s}evi\'{c}},\ and\
  \citenamefont {Mali}}]{SSC}%
  \BibitemOpen
  \bibfield  {author} {\bibinfo {author} {\bibfnamefont {M.~R.}\ \bibnamefont
  {Panti\'{c}}}, \bibinfo {author} {\bibfnamefont {D.~V.}\ \bibnamefont
  {Kapor}}, \bibinfo {author} {\bibfnamefont {S.~M.}\ \bibnamefont
  {Rado\v{s}evi\'{c}}}, \ and\ \bibinfo {author} {\bibfnamefont {P.~M.}\
  \bibnamefont {Mali}},\ }\href
  {http://www.sciencedirect.com/science/article/pii/S0038109813005656}
  {\bibfield  {journal} {\bibinfo  {journal} {Solid State Commun.}\ }\textbf
  {\bibinfo {volume} {182}},\ \bibinfo {pages} {55 } (\bibinfo {year}
  {2014})}\BibitemShut {NoStop}%
\bibitem [{\citenamefont {Hofmann}(1999)}]{Hofman1}%
  \BibitemOpen
  \bibfield  {author} {\bibinfo {author} {\bibfnamefont {C.~P.}\ \bibnamefont
  {Hofmann}},\ }\href {https://link.aps.org/doi/10.1103/PhysRevB.60.388}
  {\bibfield  {journal} {\bibinfo  {journal} {Phys. Rev. B}\ }\textbf {\bibinfo
  {volume} {60}},\ \bibinfo {pages} {388} (\bibinfo {year} {1999})}\BibitemShut
  {NoStop}%
\bibitem [{\citenamefont {Hofmann}(2012)}]{Hofman2}%
  \BibitemOpen
  \bibfield  {author} {\bibinfo {author} {\bibfnamefont {C.~P.}\ \bibnamefont
  {Hofmann}},\ }\href {https://link.aps.org/doi/10.1103/PhysRevB.86.054409}
  {\bibfield  {journal} {\bibinfo  {journal} {Phys. Rev. B}\ }\textbf {\bibinfo
  {volume} {86}},\ \bibinfo {pages} {054409} (\bibinfo {year}
  {2012})}\BibitemShut {NoStop}%
\bibitem [{\citenamefont {Rado{\v{s}}evi{\'c}}\ \emph
  {et~al.}(2013)\citenamefont {Rado{\v{s}}evi{\'c}}, \citenamefont
  {Panti{\'c}}, \citenamefont {Pavkov-Hrvojevi{\'c}},\ and\ \citenamefont
  {Kapor}}]{radosevic2013}%
  \BibitemOpen
  \bibfield  {author} {\bibinfo {author} {\bibfnamefont {S.~M.}\ \bibnamefont
  {Rado{\v{s}}evi{\'c}}}, \bibinfo {author} {\bibfnamefont {M.~R.}\
  \bibnamefont {Panti{\'c}}}, \bibinfo {author} {\bibfnamefont {M.~V.}\
  \bibnamefont {Pavkov-Hrvojevi{\'c}}}, \ and\ \bibinfo {author} {\bibfnamefont
  {D.~V.}\ \bibnamefont {Kapor}},\ }\href
  {https://www.sciencedirect.com/science/article/pii/S0003491613002054}
  {\bibfield  {journal} {\bibinfo  {journal} {Ann. Phys.}\ }\textbf {\bibinfo
  {volume} {339}},\ \bibinfo {pages} {382} (\bibinfo {year}
  {2013})}\BibitemShut {NoStop}%
\bibitem [{\citenamefont {Rado{\v{s}}evi{\'c}}(2015)}]{radosevic2015}%
  \BibitemOpen
  \bibfield  {author} {\bibinfo {author} {\bibfnamefont {S.~M.}\ \bibnamefont
  {Rado{\v{s}}evi{\'c}}},\ }\href {\doibase
  https://doi.org/10.1016/j.aop.2015.08.003} {\bibfield  {journal} {\bibinfo
  {journal} {Ann. Phys.}\ }\textbf {\bibinfo {volume} {362}},\ \bibinfo {pages}
  {336 } (\bibinfo {year} {2015})}\BibitemShut {NoStop}%
\bibitem [{\citenamefont {Stehr}\ \emph {et~al.}(2011)\citenamefont {Stehr},
  \citenamefont {Pfister}, \citenamefont {Fink}, \citenamefont {Engels},\ and\
  \citenamefont {Deibel}}]{Mob2D}%
  \BibitemOpen
  \bibfield  {author} {\bibinfo {author} {\bibfnamefont {V.}~\bibnamefont
  {Stehr}}, \bibinfo {author} {\bibfnamefont {J.}~\bibnamefont {Pfister}},
  \bibinfo {author} {\bibfnamefont {R.~F.}\ \bibnamefont {Fink}}, \bibinfo
  {author} {\bibfnamefont {B.}~\bibnamefont {Engels}}, \ and\ \bibinfo {author}
  {\bibfnamefont {C.}~\bibnamefont {Deibel}},\ }\href
  {https://link.aps.org/doi/10.1103/PhysRevB.83.155208} {\bibfield  {journal}
  {\bibinfo  {journal} {Phys. Rev. B}\ }\textbf {\bibinfo {volume} {83}},\
  \bibinfo {pages} {155208} (\bibinfo {year} {2011})}\BibitemShut {NoStop}%
\bibitem [{\citenamefont {Cornil}\ \emph {et~al.}(2001)\citenamefont {Cornil},
  \citenamefont {Calbert},\ and\ \citenamefont {Brédas}}]{Ugao}%
  \BibitemOpen
  \bibfield  {author} {\bibinfo {author} {\bibfnamefont {J.}~\bibnamefont
  {Cornil}}, \bibinfo {author} {\bibfnamefont {J.~P.}\ \bibnamefont {Calbert}},
  \ and\ \bibinfo {author} {\bibfnamefont {J.~L.}\ \bibnamefont {Brédas}},\
  }\href {https://doi.org/10.1021/ja005700i} {\bibfield  {journal} {\bibinfo
  {journal} {J. Am. Chem. Soc.}\ }\textbf {\bibinfo {volume} {123}},\ \bibinfo
  {pages} {1250} (\bibinfo {year} {2001})}\BibitemShut {NoStop}%
\bibitem [{\citenamefont {Holmes}\ \emph {et~al.}(1999)\citenamefont {Holmes},
  \citenamefont {Kumaraswamy}, \citenamefont {Matzger},\ and\ \citenamefont
  {Vollhardt}}]{Ugaotetracen}%
  \BibitemOpen
  \bibfield  {author} {\bibinfo {author} {\bibfnamefont {D.}~\bibnamefont
  {Holmes}}, \bibinfo {author} {\bibfnamefont {S.}~\bibnamefont {Kumaraswamy}},
  \bibinfo {author} {\bibfnamefont {A.~J.}\ \bibnamefont {Matzger}}, \ and\
  \bibinfo {author} {\bibfnamefont {K.~P.~C.}\ \bibnamefont {Vollhardt}},\
  }\href
  {https://onlinelibrary.wiley.com/doi/abs/10.1002/(SICI)1521-3765(19991105)5:11%3C3399::AID-CHEM3399%3E3.0.CO;2-V}
  {\bibfield  {journal} {\bibinfo  {journal} {Chem. Eur. J.}\ }\textbf
  {\bibinfo {volume} {5}},\ \bibinfo {pages} {3399} (\bibinfo {year}
  {1999})}\BibitemShut {NoStop}%
\bibitem [{\citenamefont {Roth}\ \emph
  {et~al.}(2012{\natexlab{b}})\citenamefont {Roth}, \citenamefont {Mahns},
  \citenamefont {Sch{\"o}nfelder}, \citenamefont {Hampel}, \citenamefont
  {Nohr}, \citenamefont {B{\"u}chner},\ and\ \citenamefont
  {Knupfer}}]{Ugaokrizen}%
  \BibitemOpen
  \bibfield  {author} {\bibinfo {author} {\bibfnamefont {F.}~\bibnamefont
  {Roth}}, \bibinfo {author} {\bibfnamefont {B.}~\bibnamefont {Mahns}},
  \bibinfo {author} {\bibfnamefont {R.}~\bibnamefont {Sch{\"o}nfelder}},
  \bibinfo {author} {\bibfnamefont {S.}~\bibnamefont {Hampel}}, \bibinfo
  {author} {\bibfnamefont {M.}~\bibnamefont {Nohr}}, \bibinfo {author}
  {\bibfnamefont {B.}~\bibnamefont {B{\"u}chner}}, \ and\ \bibinfo {author}
  {\bibfnamefont {M.}~\bibnamefont {Knupfer}},\ }\href
  {https://aip.scitation.org/doi/abs/10.1063/1.4753999} {\bibfield  {journal}
  {\bibinfo  {journal} {J. Chem. Phys.}\ }\textbf {\bibinfo {volume} {137}},\
  \bibinfo {pages} {114508} (\bibinfo {year} {2012}{\natexlab{b}})}\BibitemShut
  {NoStop}%
\bibitem [{\citenamefont {Amy}\ \emph {et~al.}(2005)\citenamefont {Amy},
  \citenamefont {Chan},\ and\ \citenamefont {Kahn}}]{amy}%
  \BibitemOpen
  \bibfield  {author} {\bibinfo {author} {\bibfnamefont {F.}~\bibnamefont
  {Amy}}, \bibinfo {author} {\bibfnamefont {C.}~\bibnamefont {Chan}}, \ and\
  \bibinfo {author} {\bibfnamefont {A.}~\bibnamefont {Kahn}},\ }\href
  {https://www.sciencedirect.com/science/article/pii/S156611990500011X}
  {\bibfield  {journal} {\bibinfo  {journal} {Org. Electron.}\ }\textbf
  {\bibinfo {volume} {\textbf 6}},\ \bibinfo {pages} {85} (\bibinfo {year}
  {2005})}\BibitemShut {NoStop}%
\bibitem [{\citenamefont {Sato}\ \emph {et~al.}(1981)\citenamefont {Sato},
  \citenamefont {Seki},\ and\ \citenamefont {Inokuchi}}]{sato2}%
  \BibitemOpen
  \bibfield  {author} {\bibinfo {author} {\bibfnamefont {N.}~\bibnamefont
  {Sato}}, \bibinfo {author} {\bibfnamefont {K.}~\bibnamefont {Seki}}, \ and\
  \bibinfo {author} {\bibfnamefont {H.}~\bibnamefont {Inokuchi}},\ }\href
  {https://pubs.rsc.org/en/content/articlelanding/1981/f2/f29817701621/unauth#!divAbstract}
  {\bibfield  {journal} {\bibinfo  {journal} {J. Chem. Soc. Faraday Trans.}\
  }\textbf {\bibinfo {volume} {\textbf{77}}},\ \bibinfo {pages} {1621}
  (\bibinfo {year} {1981})}\BibitemShut {NoStop}%
\bibitem [{\citenamefont {Roth}\ \emph {et~al.}(2010)\citenamefont {Roth},
  \citenamefont {Gatti}, \citenamefont {Cudazzo}, \citenamefont {Grobosch},
  \citenamefont {Mahns}, \citenamefont {B{\"u}chner}, \citenamefont {Rubio},\
  and\ \citenamefont {Knupfer}}]{picengep}%
  \BibitemOpen
  \bibfield  {author} {\bibinfo {author} {\bibfnamefont {F.}~\bibnamefont
  {Roth}}, \bibinfo {author} {\bibfnamefont {M.}~\bibnamefont {Gatti}},
  \bibinfo {author} {\bibfnamefont {P.}~\bibnamefont {Cudazzo}}, \bibinfo
  {author} {\bibfnamefont {M.}~\bibnamefont {Grobosch}}, \bibinfo {author}
  {\bibfnamefont {B.}~\bibnamefont {Mahns}}, \bibinfo {author} {\bibfnamefont
  {B.}~\bibnamefont {B{\"u}chner}}, \bibinfo {author} {\bibfnamefont
  {A.}~\bibnamefont {Rubio}}, \ and\ \bibinfo {author} {\bibfnamefont
  {M.}~\bibnamefont {Knupfer}},\ }\href
  {https://iopscience.iop.org/article/10.1088/1367-2630/12/10/103036/meta}
  {\bibfield  {journal} {\bibinfo  {journal} {New J. Phys.}\ }\textbf {\bibinfo
  {volume} {12}},\ \bibinfo {pages} {103036} (\bibinfo {year}
  {2010})}\BibitemShut {NoStop}%
\bibitem [{\citenamefont {Sato}\ \emph {et~al.}(1987)\citenamefont {Sato},
  \citenamefont {Inokuchi},\ and\ \citenamefont {Silinsh}}]{sato}%
  \BibitemOpen
  \bibfield  {author} {\bibinfo {author} {\bibfnamefont {N.}~\bibnamefont
  {Sato}}, \bibinfo {author} {\bibfnamefont {H.}~\bibnamefont {Inokuchi}}, \
  and\ \bibinfo {author} {\bibfnamefont {E.~A.}\ \bibnamefont {Silinsh}},\
  }\href {https://www.sciencedirect.com/science/article/pii/0301010487800411}
  {\bibfield  {journal} {\bibinfo  {journal} {Chem. Phys.}\ }\textbf {\bibinfo
  {volume} {\textbf {115}}},\ \bibinfo {pages} {269} (\bibinfo {year}
  {1987})}\BibitemShut {NoStop}%
\bibitem [{\citenamefont {L\"auchli}\ \emph {et~al.}(2006)\citenamefont
  {L\"auchli}, \citenamefont {Mila},\ and\ \citenamefont {Penc}}]{primerPRL}%
  \BibitemOpen
  \bibfield  {author} {\bibinfo {author} {\bibfnamefont {A.}~\bibnamefont
  {L\"auchli}}, \bibinfo {author} {\bibfnamefont {F.}~\bibnamefont {Mila}}, \
  and\ \bibinfo {author} {\bibfnamefont {K.}~\bibnamefont {Penc}},\ }\href
  {\doibase 10.1103/PhysRevLett.97.087205} {\bibfield  {journal} {\bibinfo
  {journal} {Phys. Rev. Lett.}\ }\textbf {\bibinfo {volume} {97}},\ \bibinfo
  {pages} {087205} (\bibinfo {year} {2006})}\BibitemShut {NoStop}%
\bibitem [{\citenamefont {Passell}\ \emph {et~al.}(1976)\citenamefont
  {Passell}, \citenamefont {Dietrich},\ and\ \citenamefont
  {Als-Nielsen}}]{primerEuO}%
  \BibitemOpen
  \bibfield  {author} {\bibinfo {author} {\bibfnamefont {L.}~\bibnamefont
  {Passell}}, \bibinfo {author} {\bibfnamefont {O.~W.}\ \bibnamefont
  {Dietrich}}, \ and\ \bibinfo {author} {\bibfnamefont {J.}~\bibnamefont
  {Als-Nielsen}},\ }\href {\doibase 10.1103/PhysRevB.14.4897} {\bibfield
  {journal} {\bibinfo  {journal} {Phys. Rev. B}\ }\textbf {\bibinfo {volume}
  {14}},\ \bibinfo {pages} {4897} (\bibinfo {year} {1976})}\BibitemShut
  {NoStop}%
\bibitem [{\citenamefont {Dietrich}\ \emph {et~al.}(1976)\citenamefont
  {Dietrich}, \citenamefont {Als-Nielsen},\ and\ \citenamefont
  {Passell}}]{primer2EuO}%
  \BibitemOpen
  \bibfield  {author} {\bibinfo {author} {\bibfnamefont {O.~W.}\ \bibnamefont
  {Dietrich}}, \bibinfo {author} {\bibfnamefont {J.}~\bibnamefont
  {Als-Nielsen}}, \ and\ \bibinfo {author} {\bibfnamefont {L.}~\bibnamefont
  {Passell}},\ }\href {\doibase 10.1103/PhysRevB.14.4923} {\bibfield  {journal}
  {\bibinfo  {journal} {Phys. Rev. B}\ }\textbf {\bibinfo {volume} {14}},\
  \bibinfo {pages} {4923} (\bibinfo {year} {1976})}\BibitemShut {NoStop}%
\bibitem [{\citenamefont {Rado{\v{s}}evi{\'{c}}}\ \emph
  {et~al.}(2009)\citenamefont {Rado{\v{s}}evi{\'{c}}}, \citenamefont
  {Pavkov-Hrvojevi{\'{c}}}, \citenamefont {Panti{\'{c}}}, \citenamefont
  {Rutonjski}, \citenamefont {Kapor},\ and\ \citenamefont
  {{\v{S}}krinjar}}]{primerSloba}%
  \BibitemOpen
  \bibfield  {author} {\bibinfo {author} {\bibfnamefont {S.}~\bibnamefont
  {Rado{\v{s}}evi{\'{c}}}}, \bibinfo {author} {\bibfnamefont {M.}~\bibnamefont
  {Pavkov-Hrvojevi{\'{c}}}}, \bibinfo {author} {\bibfnamefont {M.}~\bibnamefont
  {Panti{\'{c}}}}, \bibinfo {author} {\bibfnamefont {M.}~\bibnamefont
  {Rutonjski}}, \bibinfo {author} {\bibfnamefont {D.}~\bibnamefont {Kapor}}, \
  and\ \bibinfo {author} {\bibfnamefont {M.}~\bibnamefont {{\v{S}}krinjar}},\
  }\href {\doibase 10.1140/epjb/e2009-00127-2} {\bibfield  {journal} {\bibinfo
  {journal} {Eur. Phys. J. B}\ }\textbf {\bibinfo {volume} {68}},\ \bibinfo
  {pages} {511} (\bibinfo {year} {2009})}\BibitemShut {NoStop}%
\end{thebibliography}%

\end{document}